\documentclass[10pt,a4paper,onecolumn]{article}
\usepackage{marginnote}
\usepackage{graphicx}
\usepackage{xcolor}
\usepackage{authblk,etoolbox}
\usepackage{titlesec}
\usepackage{calc}
\usepackage{tikz}
\usepackage{hyperref}
\hypersetup{colorlinks,breaklinks,
            urlcolor=[rgb]{0.0, 0.5, 1.0},
            linkcolor=[rgb]{0.0, 0.5, 1.0}}
\usepackage{caption}
\usepackage{tcolorbox}
\usepackage{amssymb,amsmath}
\usepackage{ifxetex,ifluatex}
\usepackage{seqsplit}
\usepackage{xstring}

\usepackage{float}
\let\origfigure\figure
\let\endorigfigure\endfigure
\renewenvironment{figure}[1][2] {
    \expandafter\origfigure\expandafter[H]
} {
    \endorigfigure
}

\usepackage{fixltx2e} 
\usepackage[
  backend=biber,
]{biblatex}
\bibliography{paper.bib}

\let\textttOrig=\texttt
\def\texttt#1{\expandafter\textttOrig{\seqsplit{#1}}}
\renewcommand{\seqinsert}{\ifmmode
  \allowbreak
  \else\penalty6000\hspace{0pt plus 0.02em}\fi}

\makeatletter
\let\href@Orig=\href
\def\href@Urllike#1#2{\href@Orig{#1}{\begingroup
    \def\Url@String{#2}\Url@FormatString
    \endgroup}}
\def\href@Notdoi#1#2{\def\tempa{#1}\def\tempb{#2}%
  \ifx\tempa\tempb\relax\href@Urllike{#1}{#2}\else
  \href@Orig{#1}{#2}\fi}
\def\href#1#2{%
  \IfBeginWith{#1}{https://doi.org}%
  {\href@Urllike{#1}{#2}}{\href@Notdoi{#1}{#2}}}
\makeatother

\usepackage[top=3.5cm, bottom=3cm, right=1.5cm, left=1.0cm,
            headheight=2.2cm, reversemp, includemp, marginparwidth=4.5cm]{geometry}

\titleformat{\section}
  {\normalfont\sffamily\Large\bfseries}
  {}{0pt}{}
\titleformat{\subsection}
  {\normalfont\sffamily\large\bfseries}
  {}{0pt}{}
\titleformat{\subsubsection}
  {\normalfont\sffamily\bfseries}
  {}{0pt}{}
\titleformat*{\paragraph}
  {\sffamily\normalsize}

\usepackage{fancyhdr}
\makeatletter
\let\ps@plain\ps@fancy
\fancyheadoffset[L]{4.5cm}
\fancyfootoffset[L]{4.5cm}

\definecolor{linky}{rgb}{0.0, 0.5, 1.0}

\newtcolorbox{repobox}
   {colback=red, colframe=red!75!black,
     boxrule=0.5pt, arc=2pt, left=6pt, right=6pt, top=3pt, bottom=3pt}

\newcommand{\ExternalLink}{%
   \tikz[x=1.2ex, y=1.2ex, baseline=-0.05ex]{%
       \begin{scope}[x=1ex, y=1ex]
           \clip (-0.1,-0.1)
               --++ (-0, 1.2)
               --++ (0.6, 0)
               --++ (0, -0.6)
               --++ (0.6, 0)
               --++ (0, -1);
           \path[draw,
               line width = 0.5,
               rounded corners=0.5]
               (0,0) rectangle (1,1);
       \end{scope}
       \path[draw, line width = 0.5] (0.5, 0.5)
           -- (1, 1);
       \path[draw, line width = 0.5] (0.6, 1)
           -- (1, 1) -- (1, 0.6);
       }
   }

\patchcmd{\@maketitle}{center}{flushleft}{}{}
\patchcmd{\@maketitle}{center}{flushleft}{}{}
\patchcmd{\@maketitle}{\LARGE}{\LARGE\sffamily}{}{}
\def\maketitle{{%
  
  \AB@maketitle}}
\makeatletter
\renewcommand\AB@affilsepx{ \protect\Affilfont}
\renewcommand\AB@affilnote[1]{{\bfseries #1}\hspace{3pt}}
\renewcommand{\affil}[2][]%
   {\newaffiltrue\let\AB@blk@and\AB@pand
      \if\relax#1\relax\def\AB@note{\AB@thenote}\else\def\AB@note{#1}%
        \setcounter{Maxaffil}{0}\fi
        \begingroup
        \let\href=\href@Orig
        \let\texttt=\textttOrig
        \let\protect\@unexpandable@protect
        \def\thanks{\protect\thanks}\def\footnote{\protect\footnote}%
        \@temptokena=\expandafter{\AB@authors}%
        {\def\\{\protect\\\protect\Affilfont}\xdef\AB@temp{#2}}%
         \xdef\AB@authors{\the\@temptokena\AB@las\AB@au@str
         \protect\\[\affilsep]\protect\Affilfont\AB@temp}%
         \gdef\AB@las{}\gdef\AB@au@str{}%
        {\def\\{, \ignorespaces}\xdef\AB@temp{#2}}%
        \@temptokena=\expandafter{\AB@affillist}%
        \xdef\AB@affillist{\the\@temptokena \AB@affilsep
          \AB@affilnote{\AB@note}\protect\Affilfont\AB@temp}%
      \endgroup
       \let\AB@affilsep\AB@affilsepx
}
\makeatother

\renewcommand\Affilfont{\sffamily\small\mdseries}
\ifnum 0\ifxetex 1\fi\ifluatex 1\fi=0 
  \usepackage[T1]{fontenc}
  \usepackage[utf8]{inputenc}

\else 
  \ifxetex
    \usepackage{mathspec}
  \else
    \usepackage{fontspec}
  \fi
  \defaultfontfeatures{Ligatures=TeX,Scale=MatchLowercase}

\fi
\IfFileExists{upquote.sty}{\usepackage{upquote}}{}
\IfFileExists{microtype.sty}{%
\usepackage{microtype}
\UseMicrotypeSet[protrusion]{basicmath} 
}{}

\usepackage{hyperref}
\hypersetup{unicode=true,
            pdftitle={latentcor: An R Package for estimating latent correlations from mixed data types},
            pdfborder={0 0 0},
            breaklinks=true}
\usepackage{color}
\usepackage{fancyvrb}

\DefineVerbatimEnvironment{Highlighting}{Verbatim}{commandchars=\\\{\}}
\newenvironment{Shaded}{}{}
\newcommand{\KeywordTok}[1]{\textcolor[rgb]{0.00,0.44,0.13}{\textbf{#1}}}
\newcommand{\DataTypeTok}[1]{\textcolor[rgb]{0.56,0.13,0.00}{#1}}
\newcommand{\DecValTok}[1]{\textcolor[rgb]{0.25,0.63,0.44}{#1}}

\newcommand{\FloatTok}[1]{\textcolor[rgb]{0.25,0.63,0.44}{#1}}

\newcommand{\StringTok}[1]{\textcolor[rgb]{0.25,0.44,0.63}{#1}}

\newcommand{\CommentTok}[1]{\textcolor[rgb]{0.38,0.63,0.69}{\textit{#1}}}

\newcommand{\OtherTok}[1]{\textcolor[rgb]{0.00,0.44,0.13}{#1}}

\newcommand{\OperatorTok}[1]{\textcolor[rgb]{0.40,0.40,0.40}{#1}}

\newcommand{\NormalTok}[1]{#1}
\usepackage{longtable,booktabs}

\let\addcontentslineOrig=\addcontentsline
\def\addcontentsline#1#2#3{\bgroup
  \let\texttt=\textttOrig\addcontentslineOrig{#1}{#2}{#3}\egroup}
\let\markbothOrig\markboth
\def\markboth#1#2{\bgroup
  \let\texttt=\textttOrig\markbothOrig{#1}{#2}\egroup}
\let\markrightOrig\markright
\def\markright#1{\bgroup
  \let\texttt=\textttOrig\markrightOrig{#1}\egroup}

\usepackage{graphicx,grffile}
\makeatletter
\def\maxwidth{\ifdim\Gin@nat@width>\linewidth\linewidth\else\Gin@nat@width\fi}
\def\maxheight{\ifdim\Gin@nat@height>\textheight\textheight\else\Gin@nat@height\fi}
\makeatother
\setkeys{Gin}{width=\maxwidth,height=\maxheight,keepaspectratio}
\IfFileExists{parskip.sty}{%
\usepackage{parskip}
}{
\setlength{\parindent}{0pt}
\setlength{\parskip}{6pt plus 2pt minus 1pt}
}
\ifx\paragraph\undefined\else
\let\oldparagraph\paragraph
\renewcommand{\paragraph}[1]{\oldparagraph{#1}\mbox{}}
\fi
\ifx\subparagraph\undefined\else
\let\oldsubparagraph\subparagraph
\renewcommand{\subparagraph}[1]{\oldsubparagraph{#1}\mbox{}}
\fi

\title{latentcor: An R Package for estimating latent correlations from mixed
data types}

        \author[1, 2]{Mingze Huang}
          \author[3, 4, 5]{Christian L. Müller}
          \author[1]{Irina Gaynanova}
    
      \affil[1]{Department of Statistics, Texas A\& M University, College Station, TX}
      \affil[2]{Department of Economics, Texas A\& M University, College Station, TX}
      \affil[3]{Ludwig-Maximilians-Universität München, Germany}
      \affil[4]{Helmholtz Zentrum München, Germany}
      \affil[5]{Flatiron Institute, New York}
  \date{\vspace{-5ex}}

\begin{document}
\maketitle

\marginpar{
  \sffamily\small

  {\bfseries DOI:} \href{https://doi.org/}{\color{linky}{}}

  \vspace{2mm}

  {\bfseries Software}
  \begin{itemize}
    \setlength\itemsep{0em}
    \item \href{}{\color{linky}{Review}} \ExternalLink
    \item \href{https://github.com/mingzehuang/latentcor}{\color{linky}{Repository}} \ExternalLink
    \item \href{}{\color{linky}{Archive}} \ExternalLink
  \end{itemize}

  \vspace{2mm}

  {\bfseries Submitted:} 11 August 2021\\
  {\bfseries Published:} 

  \vspace{2mm}
  {\bfseries License}\\
  Authors of papers retain copyright and release the work under a Creative Commons Attribution 4.0 International License (\href{https://creativecommons.org/licenses/by/4.0/}{\color{linky}{CC BY 4.0}}).
}

\section{Summary}\label{summary}

We present \texttt{latentcor}, an R package for correlation estimation
from data with mixed variable types. Mixed variables types, including
continuous, binary, ordinal, zero-inflated, or truncated data are
routinely collected in many areas of science. Accurate estimation of
correlations among such variables is often the first critical step in
statistical analysis workflows. Pearson correlation as the default
choice is not well suited for mixed data types as the underlying
normality assumption is violated. The concept of semi-parametric latent
Gaussian copula models, on the other hand, provides a unifying way to
estimate correlations between mixed data types. The R package
\texttt{latentcor} comprises a comprehensive list of these models,
enabling the estimation of correlations between any of
continuous/binary/ternary/zero-inflated (truncated) variable types. The
underlying implementation takes advantage of a fast multi-linear
interpolation scheme with an efficient choice of interpolation grid
points, thus giving the package a small memory footprint without
compromising estimation accuracy. This makes latent correlation
estimation readily available for modern high-throughput data analysis.

\section{Statement of need}\label{statement-of-need}

No R software package is currently available that allows accurate and
fast correlation estimation from mixed variable data in a unifying
manner. The popular \texttt{cor} function within R package
\texttt{stats} (Team and others 2013), for instance, allows to compute
Pearson's correlation, Kendall's \(\tau\) and Spearman's \(\rho\), and a
faster algorithm for calculating Kendall's \(\tau\) is implemented in
the R package \texttt{pcaPP} (Croux, Filzmoser, and Fritz 2013).
Pearson's correlation is not appropriate for skewed or ordinal data, and
its use leads to invalid inference in those cases. While the rank-based
Kendall's \(\tau\) and Spearman's \(\rho\) are more robust measures of
association, the resulting values do not have correlation interpretation
and can not be used as direct substitutes in statistical methods that
require correlation as input (e.g., graphical model estimation (Yoon,
Gaynanova, and Müller 2019)). The R package \texttt{polycor} (Fox 2019)
is designed for ordinal data and allows to computes polychoric
(ordinal/ordinal) and polyserial (ordinal/continuous) correlations based
on latent Gaussian model. However, the package does not have
functionality for zero-inflated data, nor can it handle skewed
continuous measurements as it does not allow for copula transformation.
The R package \texttt{correlation} (Makowski et al. 2020) in the
\texttt{easystats} collection provides 16 different correlation
measures, including polychoric and polyserial correlations. However,
functionality for correlation estimation from zero-inflated data is
lacking. The R package \texttt{mixedCCA} (Yoon, Carroll, and Gaynanova
2020) is based on the latent Gaussian copula model and can compute
latent correlations between continuous/binary/zero-inflated variable
types as an intermediate step for canonical correlation analysis.
However, \texttt{mixedCCA} does not allow for ordinal data types. The R
package \texttt{latentcor}, introduced here, thus represents the first
stand-alone R package for computation of latent correlation that takes
into account all variable types
(continuous/binary/ordinal/zero-inflated), comes with an optimized
memory footprint, and is computationally efficient, essentially making
latent correlation estimation almost as fast as rank-based correlation
estimation.

\section{Estimation of latent
correlations}\label{estimation-of-latent-correlations}

\subsection{The general estimation
workflow}\label{the-general-estimation-workflow}

The estimation of latent correlations consists of three steps:

\begin{itemize}
\item
  computing Kendall's \(\tau\) between each pair of variables,
\item
  choosing the bridge function \(F(\cdot)\) based on the types of
  variable pairs; the bridge function connects the Kendall's \(\tau\)
  computed from the data, \(\widehat \tau\), to the true underlying
  correlation \(\rho\) via moment equation
  \(\mathbb{E}(\widehat \tau) = F(\rho)\);
\item
  estimating latent correlation by calculating
  \(F^{-1}(\widehat \tau)\).
\end{itemize}

We summarize the references for the explicit form of \(F(\cdot)\) for
each variable combination as implemented in \texttt{latentcor} below.

\begin{longtable}[]{@{}lllll@{}}
\toprule
\begin{minipage}[b]{0.14\columnwidth}\raggedright\strut
Type\strut
\end{minipage} & \begin{minipage}[b]{0.20\columnwidth}\raggedright\strut
continuous\strut
\end{minipage} & \begin{minipage}[b]{0.15\columnwidth}\raggedright\strut
binary\strut
\end{minipage} & \begin{minipage}[b]{0.22\columnwidth}\raggedright\strut
ternary\strut
\end{minipage} & \begin{minipage}[b]{0.15\columnwidth}\raggedright\strut
zero-inflated\\
(truncated)\strut
\end{minipage}\tabularnewline
\midrule
\endhead
\begin{minipage}[t]{0.14\columnwidth}\raggedright\strut
continuous\strut
\end{minipage} & \begin{minipage}[t]{0.20\columnwidth}\raggedright\strut
Liu, Lafferty, and Wasserman (2009)\strut
\end{minipage} & \begin{minipage}[t]{0.15\columnwidth}\raggedright\strut
-\strut
\end{minipage} & \begin{minipage}[t]{0.22\columnwidth}\raggedright\strut
-\strut
\end{minipage} & \begin{minipage}[t]{0.15\columnwidth}\raggedright\strut
-\strut
\end{minipage}\tabularnewline
\begin{minipage}[t]{0.14\columnwidth}\raggedright\strut
binary\strut
\end{minipage} & \begin{minipage}[t]{0.20\columnwidth}\raggedright\strut
Fan et al. (2017)\strut
\end{minipage} & \begin{minipage}[t]{0.15\columnwidth}\raggedright\strut
Fan et al. (2017)\strut
\end{minipage} & \begin{minipage}[t]{0.22\columnwidth}\raggedright\strut
-\strut
\end{minipage} & \begin{minipage}[t]{0.15\columnwidth}\raggedright\strut
-\strut
\end{minipage}\tabularnewline
\begin{minipage}[t]{0.14\columnwidth}\raggedright\strut
ternary\strut
\end{minipage} & \begin{minipage}[t]{0.20\columnwidth}\raggedright\strut
Quan, Booth, and Wells (2018)\strut
\end{minipage} & \begin{minipage}[t]{0.15\columnwidth}\raggedright\strut
Quan, Booth, and Wells (2018)\strut
\end{minipage} & \begin{minipage}[t]{0.22\columnwidth}\raggedright\strut
Quan, Booth, and Wells (2018)\strut
\end{minipage} & \begin{minipage}[t]{0.15\columnwidth}\raggedright\strut
-\strut
\end{minipage}\tabularnewline
\begin{minipage}[t]{0.14\columnwidth}\raggedright\strut
zero-inflated\\
(truncated)\strut
\end{minipage} & \begin{minipage}[t]{0.20\columnwidth}\raggedright\strut
Yoon, Carroll, and Gaynanova (2020)\strut
\end{minipage} & \begin{minipage}[t]{0.15\columnwidth}\raggedright\strut
Yoon, Carroll, and Gaynanova (2020)\strut
\end{minipage} & \begin{minipage}[t]{0.22\columnwidth}\raggedright\strut
See \texttt{latentcor}\\
vignette for derivation\\
\strut
\end{minipage} & \begin{minipage}[t]{0.15\columnwidth}\raggedright\strut
Yoon, Carroll, and Gaynanova (2020)\strut
\end{minipage}\tabularnewline
\bottomrule
\end{longtable}

\subsection{Efficient inversion of the bridge
function}\label{efficient-inversion-of-the-bridge-function}

In \texttt{latentcor}, the inversion of the bridge function \(F(\cdot)\)
can be computed in two ways. The original approach
(\texttt{method\ =\ "original"}) relies on numerical inversion for each
pair of variables based on uni-root optimization (Yoon, Carroll, and
Gaynanova 2020). Since each pair of variables requires a separate
optimization run, the original approach is computationally expensive
when the number of variables is large. The second approach to invert
\(F(\cdot)\) is through fast multi-linear interpolation of
pre-calculated \(F^{-1}\) values at specific sets of interpolation grid
points (\texttt{method\ =\ "approx"}). This construction has been
proposed in (Yoon, Müller, and Gaynanova 2021) and is available for
continuous/binary/truncated pairs in the current version of
\texttt{mixedCCA}. However, that implementation lacks the ternary
variable case and relies on an interpolation grid with a large memory
footprint. \texttt{latentcor} includes the ternary case and provides an
optimized interpolation grid by redefining the bridge functions on a
rescaled version of Kendall's \(\tau\). Here, the scaling adapts to the
smoothness of the underlying type of variables by simultaneously
controlling the approximation error at the same or lower level. As a
result, \texttt{latentcor} has significantly smaller memory footprint
(see Table below) and smaller approximation error compared to
\texttt{mixedCCA}.

Memory footprints (in KB):

\begin{longtable}[]{@{}lll@{}}
\toprule
case & mixedCCA & latentcor\tabularnewline
\midrule
\endhead
binary/continuous & 10.08 & 4.22\tabularnewline
binary/binary & 303.04 & 69.1\tabularnewline
truncated/continuous & 20.99 & 6.16\tabularnewline
truncated/binary & 907.95 & 92.25\tabularnewline
truncated/truncated & 687.68 & 84.33\tabularnewline
ternary/continuous & - & 125.83\tabularnewline
ternary/binary & - & 728.3\tabularnewline
ternary/truncated & - & 860.9\tabularnewline
ternary/ternary & - & 950.61\tabularnewline
\bottomrule
\end{longtable}

\subsection{Illustrative example}\label{illustrative-example}

\begin{figure}
\centering
\includegraphics{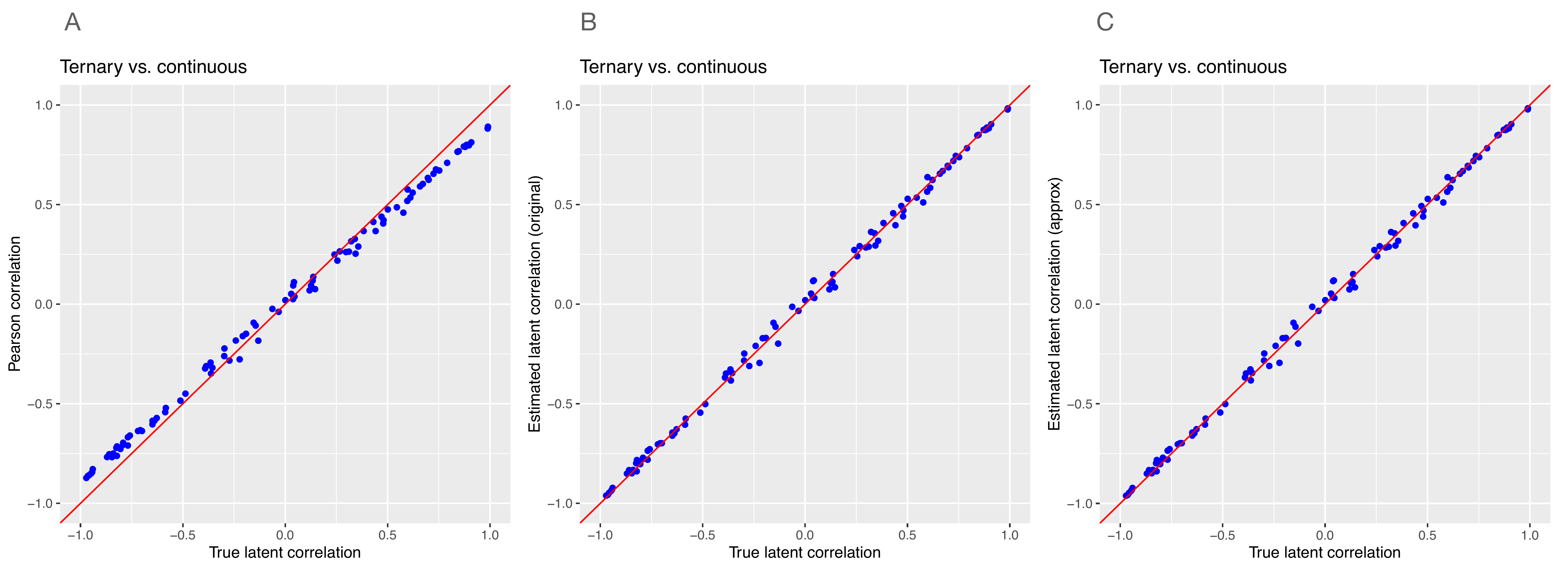}
\caption{Scatter plots of estimated Pearson correlation (panel A) and
latent correlations (\texttt{original} in panel B, \texttt{approx} in
panel C) vs.~ground truth correlations \label{fig:R_all}}
\end{figure}

To illustrate the excellent performance of latent correlation estimation
on mixed data, we consider the simple example of estimating correlations
between continuous and ternary variables. In this synthetic scenario, we
have access to the true underlying correlation between the variables.
Figure \ref{fig:R_all}A displays the values obtained by using standard
Pearson correlation, revealing a significant estimation bias with
respect to the true correlations. Figure \ref{fig:R_all}B displays the
estimated latent correlations using the original approach versus the
true values of underlying ternary/continuous correlations. The alignment
of points around \(y=x\) line confirms that the estimation is
empirically unbiased. Figure \ref{fig:R_all}C displays the estimated
latent correlations using the approximation approach
(\texttt{method\ =\ "approx"}) versus true values of underlying latent
correlation. The results are almost indistinguishable from Figure
\ref{fig:R_all}B at a fraction of the computational cost.

The script to reproduce the displayed results is available at
\href{https://github.com/mingzehuang/latentcor_evaluation/blob/master/unbias_check.R}{latentcor\_evaluation}.

\section{Basic Usage}\label{basic-usage}

We provide two basic code examples of how to use \texttt{latentcor} in
R.

The first example illustrates how to estimate latent correlation from
pairs of ternary/continuous variables.

\begin{Shaded}
\begin{Highlighting}[]
\KeywordTok{library}\NormalTok{(latentcor)}

\CommentTok{# Generate two variables of sample size 100}
\CommentTok{# The first variable is ternary (pi0 = 0.3, pi1 = 0.5, pi2 = 1-0.3-0.5 = 0.2)}
\CommentTok{# The second variable is continuous.}
\CommentTok{# No copula transformation is applied.}
\NormalTok{X =}\StringTok{ }\KeywordTok{GenData}\NormalTok{(}\DataTypeTok{types =} \KeywordTok{c}\NormalTok{(}\StringTok{"ter"}\NormalTok{, }\StringTok{"con"}\NormalTok{), }\DataTypeTok{XP =} \KeywordTok{list}\NormalTok{(}\KeywordTok{c}\NormalTok{(}\FloatTok{0.3}\NormalTok{, .}\DecValTok{5}\NormalTok{), }\OtherTok{NA}\NormalTok{))}\OperatorTok{$}\NormalTok{X}

\CommentTok{# Estimate latent correlation matrix with original method}
\KeywordTok{estR}\NormalTok{(}\DataTypeTok{X =}\NormalTok{ X, }\DataTypeTok{types =} \KeywordTok{c}\NormalTok{(}\StringTok{"ter"}\NormalTok{, }\StringTok{"con"}\NormalTok{), }\DataTypeTok{method =} \StringTok{"original"}\NormalTok{)}\OperatorTok{$}\NormalTok{R}

\CommentTok{# Estimate latent correlation matrix with approximation method}
\KeywordTok{estR}\NormalTok{(}\DataTypeTok{X =}\NormalTok{ X, }\DataTypeTok{types =} \KeywordTok{c}\NormalTok{(}\StringTok{"ter"}\NormalTok{, }\StringTok{"con"}\NormalTok{))}\OperatorTok{$}\NormalTok{R}

\CommentTok{# Heatmap for latent correlation matrix.}
\KeywordTok{estR}\NormalTok{(}\DataTypeTok{X =}\NormalTok{ X, }\DataTypeTok{types =} \KeywordTok{c}\NormalTok{(}\StringTok{"ter"}\NormalTok{, }\StringTok{"con"}\NormalTok{), }\DataTypeTok{showplot =} \OtherTok{TRUE}\NormalTok{)}\OperatorTok{$}\NormalTok{plotR}
\end{Highlighting}
\end{Shaded}

The second example considers the \texttt{mtcars} dataset, available in
standard R. The \texttt{mtcars} dataset comprises eleven variables of
continuous, binary, and ternary data type.

\begin{Shaded}
\begin{Highlighting}[]
\KeywordTok{library}\NormalTok{(latentcor)}
\CommentTok{# Use build-in dataset mtcars}
\NormalTok{X =}\StringTok{ }\NormalTok{mtcars}
\CommentTok{# Check variable types}
\KeywordTok{apply}\NormalTok{(mtcars, }\DecValTok{2}\NormalTok{, table)}
\CommentTok{# Estimate latent correlation matrix with original method}
\KeywordTok{estR}\NormalTok{(mtcars, }\DataTypeTok{types =} \KeywordTok{c}\NormalTok{(}\StringTok{"con"}\NormalTok{, }\StringTok{"ter"}\NormalTok{, }\StringTok{"con"}\NormalTok{, }\StringTok{"con"}\NormalTok{, }\StringTok{"con"}\NormalTok{, }\StringTok{"con"}\NormalTok{, }\StringTok{"con"}\NormalTok{, }\StringTok{"bin"}\NormalTok{,}
                       \StringTok{"bin"}\NormalTok{, }\StringTok{"ter"}\NormalTok{, }\StringTok{"con"}\NormalTok{), }\DataTypeTok{method =} \StringTok{"original"}\NormalTok{)}\OperatorTok{$}\NormalTok{R}
\CommentTok{# Estimate latent correlation matrix with approximation method}
\KeywordTok{estR}\NormalTok{(mtcars, }\DataTypeTok{types =} \KeywordTok{c}\NormalTok{(}\StringTok{"con"}\NormalTok{, }\StringTok{"ter"}\NormalTok{, }\StringTok{"con"}\NormalTok{, }\StringTok{"con"}\NormalTok{, }\StringTok{"con"}\NormalTok{, }\StringTok{"con"}\NormalTok{, }\StringTok{"con"}\NormalTok{, }\StringTok{"bin"}\NormalTok{,}
                       \StringTok{"bin"}\NormalTok{, }\StringTok{"ter"}\NormalTok{, }\StringTok{"con"}\NormalTok{))}\OperatorTok{$}\NormalTok{R}
\CommentTok{# Heatmap for latent correlation matrix.}
\KeywordTok{estR}\NormalTok{(mtcars, }\DataTypeTok{types =} \KeywordTok{c}\NormalTok{(}\StringTok{"con"}\NormalTok{, }\StringTok{"ter"}\NormalTok{, }\StringTok{"con"}\NormalTok{, }\StringTok{"con"}\NormalTok{, }\StringTok{"con"}\NormalTok{, }\StringTok{"con"}\NormalTok{, }\StringTok{"con"}\NormalTok{, }\StringTok{"bin"}\NormalTok{,}
                       \StringTok{"bin"}\NormalTok{, }\StringTok{"ter"}\NormalTok{, }\StringTok{"con"}\NormalTok{), }\DataTypeTok{showplot =} \OtherTok{TRUE}\NormalTok{)}\OperatorTok{$}\NormalTok{plotR}
\end{Highlighting}
\end{Shaded}

\begin{figure}
\centering
\includegraphics{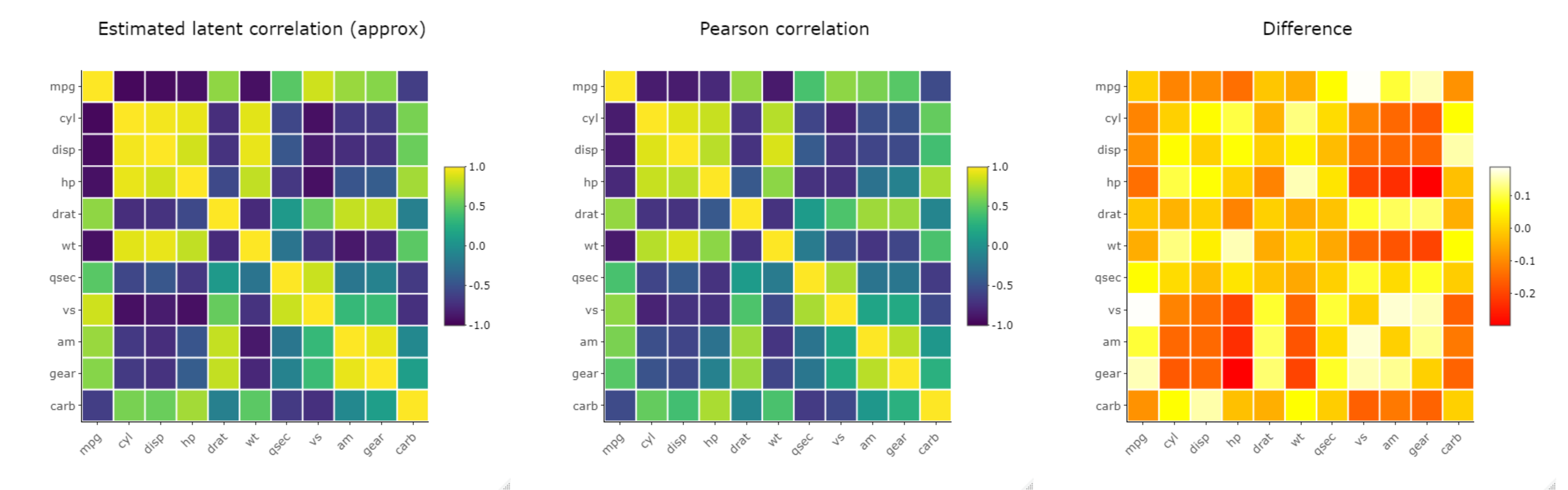}
\caption{Heatmap of latent correlations (\texttt{approx}, left panel),
Pearson correlation (middle panel), and difference between the two
estimators (latent correlation - Pearson correlation) on the mtcars
dataset \label{fig:R_cars}}
\end{figure}

Figure \ref{fig:R_cars} shows the \(11 \times 11\) matrices with latent
correlation estimates (with default \texttt{approx} method, left panel),
Pearson correlation estimates (middle panel), and their difference in
estimation (right panel). Even on this small dataset, we observe
absolute differences larger than \(0.2\).

The script to reproduce Figure \ref{fig:R_cars} is available
\href{https://github.com/mingzehuang/latentcor_evaluation/blob/master/all_heatmap.R}{here}.
We also provide interactive heatmaps for
\href{https://rpubs.com/mingzehuang/797937}{estimated latent
correlations}, \href{https://rpubs.com/mingzehuang/797945}{Pearson
correlations}, and \href{https://rpubs.com/mingzehuang/798060}{their
differences (estimated latent correlations minus Pearson correlations)}
for the \texttt{mtcars} data set.

\section{Availability}\label{availability}

The R package \texttt{latentcor} is available on
\href{https://github.com/mingzehuang/latentcor/}{Github}. A
comprehensive vignette with additional mathematical and computational
details is available
\href{https://mingzehuang.github.io/latentcor/articles/latentcor.html}{here}.

\section{Acknowledgments}\label{acknowledgments}

We thank Dr.~Grace Yoon for providing implementation details of the
\texttt{mixedCCA} R package.

\section*{References}\label{references}
\addcontentsline{toc}{section}{References}

\hypertarget{refs}{}
\hypertarget{ref-croux2013robust}{}
Croux, Christophe, Peter Filzmoser, and Heinrich Fritz. 2013. ``Robust
Sparse Principal Component Analysis.'' \emph{Technometrics} 55 (2).
Taylor \& Francis: 202--14.
doi:\href{https://doi.org/10.1080/00401706.2012.727746}{10.1080/00401706.2012.727746}.

\hypertarget{ref-fan2017high}{}
Fan, Jianqing, Han Liu, Yang Ning, and Hui Zou. 2017. ``High Dimensional
Semiparametric Latent Graphical Model for Mixed Data.'' \emph{Journal of
the Royal Statistical Society. Series B: Statistical Methodology} 79
(2). Wiley-Blackwell: 405--21.
doi:\href{https://doi.org/10.1111/rssb.12168}{10.1111/rssb.12168}.

\hypertarget{ref-fox2019poly}{}
Fox, John. 2019. \emph{Polycor: Polychoric and Polyserial Correlations}.
\url{https://CRAN.R-project.org/package=polycor}.

\hypertarget{ref-liu2009nonparanormal}{}
Liu, Han, John Lafferty, and Larry Wasserman. 2009. ``The Nonparanormal:
Semiparametric Estimation of High Dimensional Undirected Graphs.''
\emph{Journal of Machine Learning Research} 10 (10).

\hypertarget{ref-makowski2020methods}{}
Makowski, Dominique, Mattan S Ben-Shachar, Indrajeet Patil, and Daniel
Lüdecke. 2020. ``Methods and Algorithms for Correlation Analysis in R.''
\emph{Journal of Open Source Software} 5 (51): 2306.
doi:\href{https://doi.org/10.21105/joss.02306}{10.21105/joss.02306}.

\hypertarget{ref-quan2018rank}{}
Quan, Xiaoyun, James G Booth, and Martin T Wells. 2018. ``Rank-Based
Approach for Estimating Correlations in Mixed Ordinal Data.''
\emph{arXiv Preprint arXiv:1809.06255}.

\hypertarget{ref-team2013r}{}
Team, R Core, and others. 2013. ``R: A Language and Environment for
Statistical Computing.'' Vienna, Austria.

\hypertarget{ref-yoon2020sparse}{}
Yoon, Grace, Raymond J Carroll, and Irina Gaynanova. 2020. ``Sparse
Semiparametric Canonical Correlation Analysis for Data of Mixed Types.''
\emph{Biometrika} 107 (3). Oxford University Press: 609--25.
doi:\href{https://doi.org/10.1093/biomet/asaa007}{10.1093/biomet/asaa007}.

\hypertarget{ref-yoon2019microbial}{}
Yoon, Grace, Irina Gaynanova, and Christian L Müller. 2019. ``Microbial
Networks in Spring-Semi-Parametric Rank-Based Correlation and Partial
Correlation Estimation for Quantitative Microbiome Data.''
\emph{Frontiers in Genetics} 10. Frontiers: 516.
doi:\href{https://doi.org/10.3389/fgene.2019.00516}{10.3389/fgene.2019.00516}.

\hypertarget{ref-yoon2021fast}{}
Yoon, Grace, Christian L Müller, and Irina Gaynanova. 2021. ``Fast
Computation of Latent Correlations.'' \emph{Journal of Computational and
Graphical Statistics}. Taylor \& Francis, 1--8.
doi:\href{https://doi.org/10.1080/10618600.2021.1882468}{10.1080/10618600.2021.1882468}.

\end{document}